\documentclass{article}
\usepackage[utf8]{inputenc}
\usepackage{hyperref}
\usepackage{xcolor}
\usepackage[pdftex]{graphicx}
\usepackage{subfigure}
\usepackage{pbox}
\usepackage{amsmath,amssymb,amsopn,amstext,amsfonts}
\usepackage{times}
\usepackage{url}
\usepackage[space]{cite}
\usepackage{pdfsync}
\usepackage{color}
\usepackage{pdfpages}
\usepackage[normalem]{ulem}

\global\long\def\redi{\textcolor{red}{i}}

\global\long\def\bluek{\textcolor{blue}{k}}

\global\long\def\bd{\mathbf{d}}

\global\long\def\bP{\mathbf{P}}

\global\long\def\Dtau{\Delta\tau}

\global\long\def\bM{\mathbf{M}}
\global\long\def\bm{\mathbf{m}}
\global\long\def\blambda{\boldsymbol{\lambda}}
\global\long\def\bm{\mathbf{m}} % motion parameters
\title{A Tutorial on Uniform B-Spline}
\author{Yi Zhou\\
Neuromorphic Automation and Intelligence Lab (NAIL),\\
School of Robotics, Hunan University\\
Email: eeyzhou@hnu.edu.cn}
\usepackage{graphicx}
\begin{document}

\maketitle

This document facilitates understanding of core concepts about \textbf{uniform B-spline} and its matrix representation.
All the contents are borrowed from \cite{mtuBspline, qin2000general,sommer2020efficient,ding2019efficient} and rephrased such that the symbolic system and definitions are unified.

%% Q1
\section{Cox-de Boor Formula}
\label{sec:cox-de boor formula}
% There are two versions for B-spline formula.
Here we focus on the uniform case, namely all knots are evenly distributed.

A uniform B-spline of degree $k$ is defined by the control points $\bP_i~(i \in [0,N-1])$ and their corresponding weights, a.k.a the basis functions $B_{i,k}(\tau)$:
% eq
\begin{equation}
\label{eq:b-spline definition}
    \bP(\tau) \doteq \sum_{i=0}^{N-1}B_{i,k}(\tau) \bP_{i}.
\end{equation}
The number of knots are determined by $M = k + N + 1$, where $N = k + 1$.
Here we do not specify the domain on which $\bP$ is defined.
It could be either $\mathcal{R}^d$ or SE$(3)$. 

The b-spline can also be regarded as a polynomial of the temporal parameter weighted by the control points.
The polynomial of degree $k$ (i.e., the basis function at the top level) is calculated recursively from degree 0 (i.e., the bottom level).
The recursive method is called \textit{Cox-de Boor} formula, 
%% eq
\begin{align}
    B_{i,0}(\tau) &= \begin{cases} 1 &\mbox{if } \tau \in [\tau_i, \tau_{i+1}], \\
0 & \mbox{otherwise.}
\end{cases}\\
B_{i,k}(\tau) &= \frac{\tau-\tau_{i}}{\tau_{i+k} - \tau_{i}}B_{i,k-1}(\tau) + \frac{\tau_{i+k+1}-\tau}{\tau_{i+k+1} - \tau_{i+1}}B_{i+1,k-1}(\tau) \nonumber\\
&= \frac{\tau-\tau_{i}}{k\Dtau}B_{i,k-1}(\tau) + \frac{\tau_{i+k+1}-\tau}{k\Dtau}B_{i+1,k-1}(\tau),
\label{eq:cox-de boor}
\end{align}
where $\Dtau$ denotes the interval between successive knots.\\
\newpage
\noindent\textbf{How to read the Cox-de Boor formula (Eq.~\ref{eq:cox-de boor})?}

To read the formula, we need to understand the meaning of the subscripts of the basis function.
The first subscript $i$ is associated to the corresponding control point $\bP_{i}$ (which shares the same index $i$).
It is also associated to the index of the \textbf{very left knot} (i.e., $\tau_{i}$) in the corresponding non-zero domain (see the triangular computation scheme in \cite{mtuBspline}).
The second subscript $k$ denotes the degree of the basis function.
The higher the degree, the wider the non-zero domain.
In other words, the non-zero domain can be determined by the two subscripts, namely $[\tau_i, \tau_{i+k+1}]$ for $B_{i,k}$ as an example.

The \textit{Cox-de Boor} formula can be read as: The basis function of degree $k$ at position $i$ is derived from two subordinate basis functions of degree $k-1$ at position $i$ and $i+1$, respectively.
The polynomial weights can be regarded as ``linear interpolation coefficients''\footnote{Strictly speaking, this is not a linear interpolation, because the denominators of the two weights are $\tau_{i+k}-\tau_{i}$ and $\tau_{i+k+1}-\tau_{i+1}$, respectively, though being equal numerically.} normalized by the width of the corresponding non-zero domain, which can be calculated as the \textbf{$\mathbf{2_{\text{nd}}}$ subscript $\mathbf{+ 1 +}$} the \textbf{$\mathbf{1_{\text{st}}}$ subscript $\mathbf{-}$} the \textbf{$\mathbf{1_{\text{st}}}$ subscript}, namely the \textbf{$2_{\text{nd}}$ subscript $\mathbf{+ 1}$}.
In other words, the width of the non-zero domain for $B_{i,k}$ is $(k+1)\Dtau$.
%+++++++++++++++++++++++++++++++++++++++++++++++++++++++++++++++++++++++++++
\subsection{Cumulative Formula}
Eq.~\ref{eq:b-spline definition} can also be represented by the cumulative form,
% eq
\begin{align}
\label{eq:cumulative formula}
    \bP(\tau) &= \tilde{B}_{0,k}(\tau)\bP_0 + \sum_{i=1}^{N-1} \tilde{B}_{i,k}(\tau)(\bP_i-\bP_{i-1}),\\
    \tilde{B}_{i,k}(\tau) &= \sum_{s=i}^{N-1}B_{s,k}(\tau).\nonumber
\label{eq:cumulative formula}
\end{align}

\section{Matrix Representation of the Cox-de Boor formula}

B-splines have local support, which means that for a spline of degree $k$, only $k+1$ control points contribute to the value of the spline at a given $\tau$.
As shown in \cite{qin2000general}, it is possible to represent the spline coefficients using a matrix representation, which is constant for uniform B-splines.

% why do we need matrix representation
An explicitly recursive matrix formula was presented in \cite{qin2000general} for non-uniform B-spline curves of an arbitrary degree by means of the Toeplitz matrix.
% Specifically, new matrix formulas for uniform B-splines and Bézier curves are given.
% \textcolor{red}{Rephrase this paragraph.}
In this section, we first revisit the idea of the Toeplitz matrix, based on which the matrix representation of the Cox-de Boor formula is derived.

% As the special case to the non-uniform configuration, the matrix formulas for uniform B-splines and Bézier curves can be derived similarly.

%+++++++++++++++++++++++++++++++++++++++++++++++++++++
\subsection{Toeplitz Matrix}
The Toeplitz matrix is a banded-shape matrix, whose elements on any line parallel to the main diagonal are all equal.
A special Toeplitz matrix is a lower triangular matrix

%eq
\begin{equation}
\mathbf{T} = 
\begin{bmatrix}
a_0 & 0 & 0 & 0 & 0 & 0 \\
a_1 & a_0 & 0 & 0 & 0 & 0 \\
\vdots & \ddots & \ddots & 0 & 0 & 0 \\
a_n & \hdots & \ddots & \ddots & 0 & 0 \\
0 & a_n & \hdots & \ddots & \ddots & 0 \\
0 & 0 & a_n & \hdots & a_1 & a_0 
\end{bmatrix},
\end{equation}
whose elements are the coefficients of the following polynomial,
%eq
\begin{equation}
    f(x) = a_0 + a_1 x + \hdots + a_n x^n (n \neq 0).
\end{equation}

\noindent ($\star$) \textbf{Toeplitz matrix can also be used to represent the product of two polynomials.}\\
Here is a specific example.
Let
%eq
\begin{align}
    g(x) &= c_0 + c_1 x + \hdots + c_2 x^2, \nonumber\\
    q(x) &= d_0 + d_1 x + \hdots + d_3 x^3. \nonumber
\end{align}
One can obtain the product $f(x) = g(x)q(x)$ in the matrix representation as,
% eq
\begin{align}
f(x) &= \mathbf{X}
\begin{bmatrix}
c_0 & 0 & 0 & 0 & 0 & 0 \\
c_1 & c_0 & 0 & 0 & 0 & 0 \\
c_2 & c_1 & c_0 & 0 & 0 & 0 \\
0 & c_2 & c_1 & c_0 & 0 & 0 \\
0 & 0 & c_2 & c_1 & c_0 & 0 \\
0 & 0 & 0 & c_2 & c_1 & c_0 
\end{bmatrix}
\begin{bmatrix}
d_0 \\
d_1 \\
d_2 \\
d_3 \\
0 \\
0 
\end{bmatrix}\\
&= \mathbf{X} 
\begin{bmatrix}
c_0 & 0 & 0 & 0 \\
c_1 & c_0 & 0 & 0\\
c_2 & c_1 & c_0 & 0\\
0 & c_2 & c_1 & c_0\\
0 & 0 & c_2 & c_1\\
0 & 0 & 0 & c_2
\end{bmatrix}
\begin{bmatrix}
d_0 \\
d_1 \\
d_2 \\
d_3
\end{bmatrix},
\nonumber
\end{align}
where $\mathbf{X} = [1,x,x^2,\cdots,x^5]$.
Note that the dimension (row) of the coefficient matrix is defined by the degree of the variable (i.e., 5 + 1 = 6).
%+++++++++++++++++++++++++++++++++++++++++++++++++++++++
\subsection{Representing the Cox-de Boor Formula Using Toeplitz Matrix}
To preserve numerical stability, it is typical to 
use a normalized variable $u$, which can be transferred from $\tau$ by means of basis translation~\cite{de1972calculating}.
Thus, the basis function $B_{i,k}(u)$ can be represented as
% eq
\begin{equation}
    B_{\redi, \bluek}(u) = [1~u~u^{\textcolor{green}{2}}~\cdots~u^{\textcolor{green}{k}}]
    \begin{bmatrix}
    N_{\redi,\bluek}^{\textcolor{green}{0}} \\
    N_{\redi,\bluek}^{\textcolor{green}{1}} \\
    N_{\redi,\bluek}^{\textcolor{green}{2}} \\
    \vdots \\
    N_{\redi,\bluek}^{\textcolor{green}{k}}
\end{bmatrix},
\label{eq:basis function translated}
\end{equation}
where $N_{\{\cdot,\cdot\}}^{\{\cdot\}}$ denotes the coefficients of the polynomial. 
% (\sout{i.e., the new basis function with the normalized variable $u$ as parameter}).
The colors of the super-/sub-scripts specify the association.
Note that the superscript of $N_{\{\cdot,\cdot\}}^{\{\cdot\}}$ is only a symbol that specifies the association to the power of variable $u$ rather than a power.

The following is the derivation of the basis translation originating from Eq.~\ref{eq:cox-de boor}.

% eq
\begin{align}
    B_{i,k}(\tau) &= \frac{\tau-\tau_{i}}{\tau_{i+k} - \tau_{i}}B_{i,k-1}(\tau) + \frac{\tau_{i+k+1}-\tau}{\tau_{i+k+1} - \tau_{i+1}}B_{i+1,k-1}(\tau) \nonumber\\
    &= \frac{(\tau_{j+1} - \tau_{j})(\tau - \tau_j + \tau_j - \tau_i)}{(\tau_{j+1} - \tau_{j})(\tau_{i+k} - \tau_{i})}B_{i,k-1}(\tau) \nonumber\\
    &+ \frac{(\tau_{j+1} - \tau_{j})(\tau_{i+k+1}-\tau_{j}+\tau_{j}-\tau)}{(\tau_{j+1}-\tau_j)(\tau_{i+k+1}-\tau_{i+1})}
    B_{i+1,k-1}(\tau)\nonumber\\
    &= \bigg[\frac{\tau_{j}-\tau_{i}}{\tau_{i+k}-\tau_{i}} + \frac{\tau-\tau_j}{\tau_{j+1}-\tau_{j}}
    \frac{\tau_{j+1}-\tau_j}{\tau_{i+k}-\tau_i}\bigg]
    B_{i,k-1}(\tau)\nonumber\\
    &+ \bigg[\frac{\tau_{i+k+1}-\tau_{j}}{\tau_{i+k+1}-\tau_{i+1}} - \frac{\tau-\tau_{j}}{\tau_{j+1}-\tau_j}
    \frac{\tau_{j+1}-\tau_j}{\tau_{i+k+1} - \tau_{i+1}}
    \bigg]
    B_{i+1,k-1}(\tau),
\label{eq:basis translation}
\end{align}
where $\tau \in [\tau_j, \tau_{j+1}]$.
\textbf{In a specific case ($k=3$), the non-zero domain is $[\tau_3, \tau_4]$ (namely j = 3), and $i = 0, 1, \cdots, 3$.}
Let
%eq
\begin{align}
    u &= \frac{\tau-\tau_j}{\tau_{j+1} - \tau_{j}},\label{eq: basis translation}\\
    d_i^0 &= \frac{\tau_j - \tau_i}{\tau_{i+k}-\tau_{i}},~
    d_i^1 = \frac{\tau_{j+1}-\tau_j}{\tau_{i+k} - \tau_{i}},\\
    h_i^0 &= \frac{\tau_{i+k+1}-\tau_j}{\tau_{i+k+1}-\tau_{i+1}},~
    h_i^1 = -\frac{\tau_{j+1}-\tau_j}{\tau_{i+k+1}-\tau_{i+1}},
\end{align}
with the convention $\frac{0}{0} = 0$.
Then Eq.~\ref{eq:basis translation} turns to
%eq
\begin{equation}
    B_{i,k}(u) = (d_i^0 + u\,d_i^1)B_{i,k-1}(u)+
    (h_i^0 + u\,h_i^1)B_{i+1,k-1}(u). 
    \label{eq:cox-deboor basis translation}
\end{equation}
Using property ($\star$), Eq.~\ref{eq:cox-deboor basis translation} can be represented by a matrix.
Here, for simplicity, we use a specific case ($k=3$) as an example,
%eq
\begin{align}
\label{eq:basis function in Toeplitz representation}
    B_{i,\textcolor{blue}{3}} &= [1~u~u^{\textcolor{green}{2}}~u^{\textcolor{green}{3}}]
    \Bigg\{
    \begin{bmatrix}
    N_{i,\textcolor{blue}{2}}^{\textcolor{green}{0}} & 0 & \vert & 0 & 0 \\
    N_{i,\textcolor{blue}{2}}^{\textcolor{green}{1}} & N_{i,\textcolor{blue}{2}}^{\textcolor{green}{0}} & \vert & 0 & 0\\
    N_{i,\textcolor{blue}{2}}^{\textcolor{green}{2}} & N_{i,\textcolor{blue}{2}}^{\textcolor{green}{1}} & \vert & N_{i,\textcolor{blue}{2}}^{\textcolor{green}{0}} & 0\\
    0 & N_{i,\textcolor{blue}{2}}^{\textcolor{green}{2}} & \vert & N_{i,\textcolor{blue}{2}}^{\textcolor{green}{1}} & N_{i,\textcolor{blue}{2}}^{\textcolor{green}{0}}
    \end{bmatrix}
    \begin{bmatrix}
    d_i^{\textcolor{green}{0}}\\
    d_i^{\textcolor{green}{1}}\\
    -\\
    0\\
    0
    \end{bmatrix} \nonumber\\
    &+
    \begin{bmatrix}
    N_{i+1,\textcolor{blue}{2}}^{\textcolor{green}{0}} & 0 & \vert & 0 & 0\\
    N_{i+1,\textcolor{blue}{2}}^{\textcolor{green}{1}} & N_{i+1,\textcolor{blue}{2}}^{\textcolor{green}{0}} & \vert & 0 & 0\\
    N_{i+1,\textcolor{blue}{2}}^{\textcolor{green}{2}} & N_{i+1,\textcolor{blue}{2}}^{\textcolor{green}{1}} & \vert & N_{i+1,\textcolor{blue}{2}}^{\textcolor{green}{0}} & 0\\
    0 & N_{i+1,\textcolor{blue}{2}}^{\textcolor{green}{2}} & \vert & N_{i+1,\textcolor{blue}{2}}^{\textcolor{green}{1}} & N_{i+1,\textcolor{blue}{2}}^{\textcolor{green}{0}}
    \end{bmatrix}
    \begin{bmatrix}
    h_i^{\textcolor{green}{0}}\\
    h_i^{\textcolor{green}{1}}\\
    -\\
    0\\
    0
    \end{bmatrix}
    \Bigg\},
\end{align}
where $N_{\{\cdot,\cdot\}}^{\{\cdot\}}$ refers to the coefficients of polynomial $B_{i,k}$, and their superscripts still do not represent a power.

% what is the matrix representation for the cumulative form
\section{Representing B-Spline Curves with Basis Matrices}

\subsection{General Matrices for NURBS}
Based on the basis translation introduced in Eq.~\ref{eq:basis function translated}, the B-spline formula (Eq.~\ref{eq:b-spline definition}) can be represented as
%eq
\begin{equation}
\label{eq:basis translated b-spline definition}
    \bP(u) = \sum_{i=0}B_{i,k}(u) \bP_{i}.
\end{equation}
Still, we use $k = \textcolor{blue}{3}$ as a specific example, and therefore, we can obtain
% eq
\begin{align}
\label{eq:B-spline in basis translated representation}
    {\bP(u)}^T = [B_{\textcolor{red}{0},\textcolor{blue}{3}}(u)~B_{\textcolor{red}{1},\textcolor{blue}{3}}(u)~B_{\textcolor{red}{2},\textcolor{blue}{3}}(u)~B_{\textcolor{red}{3},\textcolor{blue}{3}}(u)]
    \begin{bmatrix}
    \bP_{\textcolor{red}{0}}^T\\
    \bP_{\textcolor{red}{1}}^T\\
    \bP_{\textcolor{red}{2}}^T\\
    \bP_{\textcolor{red}{3}}^T
    \end{bmatrix}\nonumber\\
    \stackrel{(Eq.~\ref{eq:basis function translated})}{=}
    [1~u~u^{\textcolor{green}{2}}~u^{\textcolor{green}{3}}]
    \underbrace{
    \begin{bmatrix}
    N_{\textcolor{red}{0},\textcolor{blue}{3}}^{\textcolor{green}{0}} & N_{\textcolor{red}{1},\textcolor{blue}{3}}^{\textcolor{green}{0}} & N_{\textcolor{red}{2},\textcolor{blue}{3}}^{\textcolor{green}{0}} & N_{\textcolor{red}{3},\textcolor{blue}{3}}^{\textcolor{green}{0}} \\
    N_{\textcolor{red}{0},\textcolor{blue}{3}}^{\textcolor{green}{1}} & N_{\textcolor{red}{1},\textcolor{blue}{3}}^{\textcolor{green}{1}} & N_{\textcolor{red}{2},\textcolor{blue}{3}}^{\textcolor{green}{1}} & N_{\textcolor{red}{3},\textcolor{blue}{3}}^{\textcolor{green}{1}} \\
    N_{\textcolor{red}{0},\textcolor{blue}{3}}^{\textcolor{green}{2}} & N_{\textcolor{red}{1},\textcolor{blue}{3}}^{\textcolor{green}{2}} & N_{\textcolor{red}{2},\textcolor{blue}{3}}^{\textcolor{green}{2}} & N_{\textcolor{red}{3},\textcolor{blue}{3}}^{\textcolor{green}{2}} \\
    N_{\textcolor{red}{0},\textcolor{blue}{3}}^{\textcolor{green}{3}} & N_{\textcolor{red}{1},\textcolor{blue}{3}}^{\textcolor{green}{3}} & N_{\textcolor{red}{2},\textcolor{blue}{3}}^{\textcolor{green}{3}} & N_{\textcolor{red}{3},\textcolor{blue}{3}}^{\textcolor{green}{3}} \\
    \end{bmatrix}
    }_{\mathbf{M}^{\textcolor{blue}{3}}(3)}
    \begin{bmatrix}
    \bP_{\textcolor{red}{0}}^T\\
    \bP_{\textcolor{red}{1}}^T\\
    \bP_{\textcolor{red}{2}}^T\\
    \bP_{\textcolor{red}{3}}^T
    \end{bmatrix},
\end{align}
where $u = \frac{\tau-\tau_3}{\tau_4-\tau_3} \in [0,1]$.
The matrix $\mathbf{M}^{k}(j)$ is referred to as \textbf{basis matrix}.
The core of this section is to derive the recursive formula for the basis matrices of B-splines of degree $k$.

\begin{figure}[h]
\vspace{2ex}
% \captionsetup{skip=3ex}
  \centering
  \includegraphics[width=0.95\textwidth]{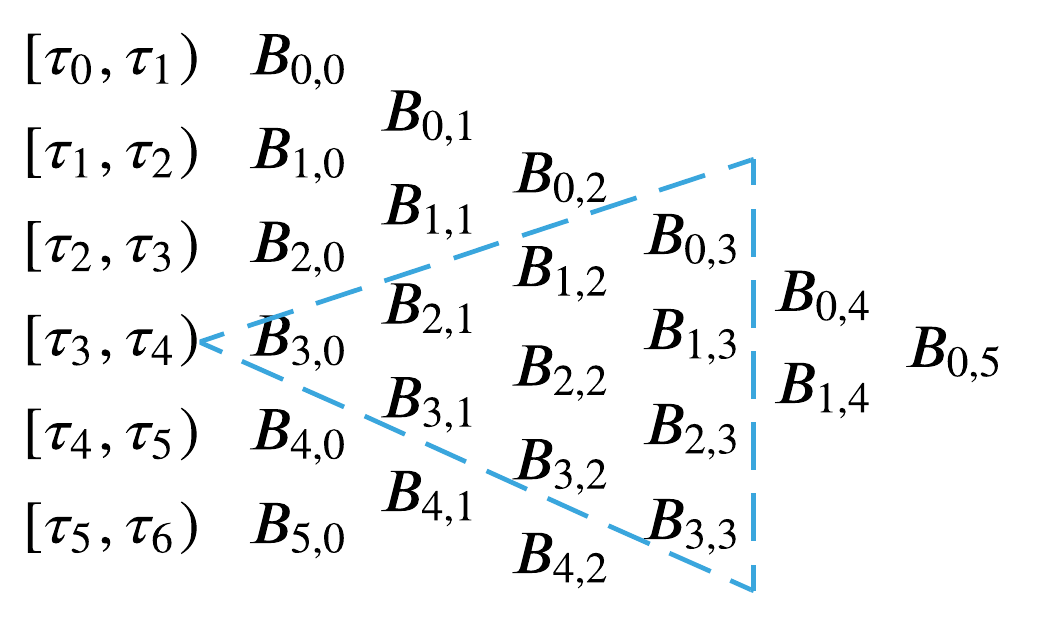}
  \caption{The triangular computation scheme of B-spline.}
  \label{fig:B-spline Triangle Computation Scheme}
  %\vspace{-2ex}
\end{figure}

According to Eq.~\ref{eq:basis function in Toeplitz representation}, the basis matrix $\mathbf{M}^{\textcolor{blue}{3}}(3)$ can be represented as
% eq
\begin{align}
\label{eq:basis matrix in Toeplitz representation}
\bM^{\textcolor{blue}{3}}(3) &= 
% line 1
\begin{bmatrix}
    N_{\textcolor{red}{0},\textcolor{blue}{3}}^{\textcolor{green}{0}} & 0 & 0 & 0\\
    N_{\textcolor{red}{0},\textcolor{blue}{3}}^{\textcolor{green}{1}} & 0 & 0 & 0\\
    N_{\textcolor{red}{0},\textcolor{blue}{3}}^{\textcolor{green}{2}} & 0 & 0 & 0\\
    N_{\textcolor{red}{0},\textcolor{blue}{3}}^{\textcolor{green}{3}} & 0 & 0 & 0
\end{bmatrix}
+
\begin{bmatrix}
    0 & N_{\textcolor{red}{1},\textcolor{blue}{3}}^{\textcolor{green}{0}} & 0 & 0\\
    0 & N_{\textcolor{red}{1},\textcolor{blue}{3}}^{\textcolor{green}{1}} & 0 & 0\\
    0 & N_{\textcolor{red}{1},\textcolor{blue}{3}}^{\textcolor{green}{2}} & 0 & 0\\
    0 & N_{\textcolor{red}{1},\textcolor{blue}{3}}^{\textcolor{green}{3}} & 0 & 0
\end{bmatrix}\nonumber\\
&+
\begin{bmatrix}
    0 & 0 & N_{\textcolor{red}{2},\textcolor{blue}{3}}^{\textcolor{green}{0}} & 0\\
    0 & 0 & N_{\textcolor{red}{2},\textcolor{blue}{3}}^{\textcolor{green}{1}} & 0\\
    0 & 0 & N_{\textcolor{red}{2},\textcolor{blue}{3}}^{\textcolor{green}{2}} & 0\\
    0 & 0 & N_{\textcolor{red}{2},\textcolor{blue}{3}}^{\textcolor{green}{3}} & 0
\end{bmatrix}
+
\begin{bmatrix}
    0 & 0 & 0 & N_{\textcolor{red}{3},\textcolor{blue}{3}}^{\textcolor{green}{0}}\\
    0 & 0 & 0 & N_{\textcolor{red}{3},\textcolor{blue}{3}}^{\textcolor{green}{1}}\\
    0 & 0 & 0 & N_{\textcolor{red}{3},\textcolor{blue}{3}}^{\textcolor{green}{2}}\\
    0 & 0 & 0 & N_{\textcolor{red}{3},\textcolor{blue}{3}}^{\textcolor{green}{3}}
\end{bmatrix}\nonumber\\
&= 
% line 2
\begin{bmatrix}
    N_{\textcolor{red}{0},\textcolor{blue}{2}}^{\textcolor{green}{0}} & 0\\
    N_{\textcolor{red}{0},\textcolor{blue}{2}}^{\textcolor{green}{1}} & N_{\textcolor{red}{0},\textcolor{blue}{2}}^{\textcolor{green}{0}}\\
    N_{\textcolor{red}{0},\textcolor{blue}{2}}^{\textcolor{green}{2}} & N_{\textcolor{red}{0},\textcolor{blue}{2}}^{\textcolor{green}{1}}\\
    0 & N_{\textcolor{red}{0},\textcolor{blue}{2}}^{\textcolor{green}{2}}
\end{bmatrix}
\begin{bmatrix}
    d_{\textcolor{red}{0}}^{\textcolor{green}{0}} & 0 & 0 & 0\\
    d_{\textcolor{red}{0}}^{\textcolor{green}{1}} & 0 & 0 & 0
\end{bmatrix}
+
\begin{bmatrix}
    N_{\textcolor{red}{1},\textcolor{blue}{2}}^{\textcolor{green}{0}} & 0\\
    N_{\textcolor{red}{1},\textcolor{blue}{2}}^{\textcolor{green}{1}} & N_{\textcolor{red}{1},\textcolor{blue}{2}}^{\textcolor{green}{0}}\\
    N_{\textcolor{red}{1},\textcolor{blue}{2}}^{\textcolor{green}{2}} & N_{\textcolor{red}{1},\textcolor{blue}{2}}^{\textcolor{green}{1}}\\
    0 & N_{\textcolor{red}{1},\textcolor{blue}{2}}^{\textcolor{green}{2}}
\end{bmatrix}
\begin{bmatrix}
    h_{\textcolor{red}{0}}^{\textcolor{green}{0}} & 0 & 0 & 0\\
    h_{\textcolor{red}{0}}^{\textcolor{green}{1}} & 0 & 0 & 0
\end{bmatrix}\nonumber\\
&+
\begin{bmatrix}
    N_{\textcolor{red}{1},\textcolor{blue}{2}}^{\textcolor{green}{0}} & 0\\
    N_{\textcolor{red}{1},\textcolor{blue}{2}}^{\textcolor{green}{1}} & N_{\textcolor{red}{1},\textcolor{blue}{2}}^{\textcolor{green}{0}}\\
    N_{\textcolor{red}{1},\textcolor{blue}{2}}^{\textcolor{green}{2}} & N_{\textcolor{red}{1},\textcolor{blue}{2}}^{\textcolor{green}{1}}\\
    0 & N_{\textcolor{red}{1},\textcolor{blue}{2}}^{\textcolor{green}{2}}
\end{bmatrix}
\begin{bmatrix}
    0 & d_{\textcolor{red}{1}}^{\textcolor{green}{0}} & 0 & 0\\
    0 & d_{\textcolor{red}{1}}^{\textcolor{green}{1}} & 0 & 0
\end{bmatrix}
+
\begin{bmatrix}
    N_{\textcolor{red}{2},\textcolor{blue}{2}}^{\textcolor{green}{0}} & 0\\
    N_{\textcolor{red}{2},\textcolor{blue}{2}}^{\textcolor{green}{1}} & N_{\textcolor{red}{2},\textcolor{blue}{2}}^{\textcolor{green}{0}}\\
    N_{\textcolor{red}{2},\textcolor{blue}{2}}^{\textcolor{green}{2}} & N_{\textcolor{red}{2},\textcolor{blue}{2}}^{\textcolor{green}{1}}\\
    0 & N_{\textcolor{red}{2},\textcolor{blue}{2}}^{\textcolor{green}{2}}
\end{bmatrix}
\begin{bmatrix}
    0 & h_{\textcolor{red}{1}}^{\textcolor{green}{0}} & 0 & 0\\
    0 & h_{\textcolor{red}{1}}^{\textcolor{green}{1}} & 0 & 0
\end{bmatrix}\nonumber\\
&+
\begin{bmatrix}
    N_{\textcolor{red}{2},\textcolor{blue}{2}}^{\textcolor{green}{0}} & 0\\
    N_{\textcolor{red}{2},\textcolor{blue}{2}}^{\textcolor{green}{1}} & N_{\textcolor{red}{2},\textcolor{blue}{2}}^{\textcolor{green}{0}}\\
    N_{\textcolor{red}{2},\textcolor{blue}{2}}^{\textcolor{green}{2}} & N_{\textcolor{red}{2},\textcolor{blue}{2}}^{\textcolor{green}{1}}\\
    0 & N_{\textcolor{red}{2},\textcolor{blue}{2}}^{\textcolor{green}{2}}
\end{bmatrix}
\begin{bmatrix}
    0 & 0 & d_{\textcolor{red}{2}}^{\textcolor{green}{0}} & 0\\
    0 & 0 & d_{\textcolor{red}{2}}^{\textcolor{green}{1}} & 0
\end{bmatrix}
+
\begin{bmatrix}
    N_{\textcolor{red}{3},\textcolor{blue}{2}}^{\textcolor{green}{0}} & 0\\
    N_{\textcolor{red}{3},\textcolor{blue}{2}}^{\textcolor{green}{1}} & N_{\textcolor{red}{3},\textcolor{blue}{2}}^{\textcolor{green}{0}}\\
    N_{\textcolor{red}{3},\textcolor{blue}{2}}^{\textcolor{green}{2}} & N_{\textcolor{red}{3},\textcolor{blue}{2}}^{\textcolor{green}{1}}\\
    0 & N_{\textcolor{red}{3},\textcolor{blue}{2}}^{\textcolor{green}{2}}
\end{bmatrix}
\begin{bmatrix}
    0 & 0 & h_{\textcolor{red}{2}}^{\textcolor{green}{0}} & 0\\
    0 & 0 & h_{\textcolor{red}{2}}^{\textcolor{green}{1}} & 0
\end{bmatrix}\nonumber\\
&+
\begin{bmatrix}
    N_{\textcolor{red}{3},\textcolor{blue}{2}}^{\textcolor{green}{0}} & 0\\
    N_{\textcolor{red}{3},\textcolor{blue}{2}}^{\textcolor{green}{1}} & N_{\textcolor{red}{3},\textcolor{blue}{2}}^{\textcolor{green}{0}}\\
    N_{\textcolor{red}{3},\textcolor{blue}{2}}^{\textcolor{green}{2}} & N_{\textcolor{red}{3},\textcolor{blue}{2}}^{\textcolor{green}{1}}\\
    0 & N_{\textcolor{red}{3},\textcolor{blue}{2}}^{\textcolor{green}{2}}
\end{bmatrix}
\begin{bmatrix}
    0 & 0 & 0 & d_{\textcolor{red}{3}}^{\textcolor{green}{0}}\\
    0 & 0 & 0 & d_{\textcolor{red}{3}}^{\textcolor{green}{1}}
\end{bmatrix}
+
\begin{bmatrix}
    N_{\textcolor{red}{4},\textcolor{blue}{2}}^{\textcolor{green}{0}} & 0\\
    N_{\textcolor{red}{4},\textcolor{blue}{2}}^{\textcolor{green}{1}} & N_{\textcolor{red}{4},\textcolor{blue}{2}}^{\textcolor{green}{0}}\\
    N_{\textcolor{red}{4},\textcolor{blue}{2}}^{\textcolor{green}{2}} & N_{\textcolor{red}{4},\textcolor{blue}{2}}^{\textcolor{green}{1}}\\
    0 & N_{\textcolor{red}{4},\textcolor{blue}{2}}^{\textcolor{green}{2}}
\end{bmatrix}
\begin{bmatrix}
    0 & 0 & 0 & h_{\textcolor{red}{3}}^{\textcolor{green}{0}}\\
    0 & 0 & 0 & h_{\textcolor{red}{3}}^{\textcolor{green}{1}}
\end{bmatrix}\nonumber\\
&=
%line 3
\underbrace{
\begin{bmatrix}
    N_{\textcolor{red}{0},\textcolor{blue}{2}}^{\textcolor{green}{0}} & 0\\
    N_{\textcolor{red}{0},\textcolor{blue}{2}}^{\textcolor{green}{1}} & N_{\textcolor{red}{0},\textcolor{blue}{2}}^{\textcolor{green}{0}}\\
    N_{\textcolor{red}{0},\textcolor{blue}{2}}^{\textcolor{green}{2}} & N_{\textcolor{red}{0},\textcolor{blue}{2}}^{\textcolor{green}{1}}\\
    0 & N_{\textcolor{red}{0},\textcolor{blue}{2}}^{\textcolor{green}{2}}
\end{bmatrix}}_{=\mathbf{0}}
\begin{bmatrix}
    d_{\textcolor{red}{0}}^{\textcolor{green}{0}} & 0 & 0 & 0\\
    d_{\textcolor{red}{0}}^{\textcolor{green}{1}} & 0 & 0 & 0
\end{bmatrix}
+
\begin{bmatrix}
    N_{\textcolor{red}{1},\textcolor{blue}{2}}^{\textcolor{green}{0}} & 0\\
    N_{\textcolor{red}{1},\textcolor{blue}{2}}^{\textcolor{green}{1}} & N_{\textcolor{red}{1},\textcolor{blue}{2}}^{\textcolor{green}{0}}\\
    N_{\textcolor{red}{1},\textcolor{blue}{2}}^{\textcolor{green}{2}} & N_{\textcolor{red}{1},\textcolor{blue}{2}}^{\textcolor{green}{1}}\\
    0 & N_{\textcolor{red}{1},\textcolor{blue}{2}}^{\textcolor{green}{2}}
\end{bmatrix}
\begin{bmatrix}
    h_{\textcolor{red}{0}}^{\textcolor{green}{0}} & d_{\textcolor{red}{1}}^{\textcolor{green}{0}} & 0 & 0\\
    h_{\textcolor{red}{0}}^{\textcolor{green}{1}} & d_{\textcolor{red}{1}}^{\textcolor{green}{1}} & 0 & 0
\end{bmatrix}\nonumber\\
&+
\begin{bmatrix}
    N_{\textcolor{red}{2},\textcolor{blue}{2}}^{\textcolor{green}{0}} & 0\\
    N_{\textcolor{red}{2},\textcolor{blue}{2}}^{\textcolor{green}{1}} & N_{\textcolor{red}{2},\textcolor{blue}{2}}^{\textcolor{green}{0}}\\
    N_{\textcolor{red}{2},\textcolor{blue}{2}}^{\textcolor{green}{2}} & N_{\textcolor{red}{2},\textcolor{blue}{2}}^{\textcolor{green}{1}}\\
    0 & N_{\textcolor{red}{2},\textcolor{blue}{2}}^{\textcolor{green}{2}}
\end{bmatrix}
\begin{bmatrix}
    0 & h_{\textcolor{red}{1}}^{\textcolor{green}{0}} & d_{\textcolor{red}{2}}^{\textcolor{green}{0}} & 0\\
    0 & h_{\textcolor{red}{1}}^{\textcolor{green}{1}} & d_{\textcolor{red}{2}}^{\textcolor{green}{1}} & 0
\end{bmatrix}
+
\begin{bmatrix}
    N_{\textcolor{red}{3},\textcolor{blue}{2}}^{\textcolor{green}{0}} & 0\\
    N_{\textcolor{red}{3},\textcolor{blue}{2}}^{\textcolor{green}{1}} & N_{\textcolor{red}{3},\textcolor{blue}{2}}^{\textcolor{green}{0}}\\
    N_{\textcolor{red}{3},\textcolor{blue}{2}}^{\textcolor{green}{2}} & N_{\textcolor{red}{3},\textcolor{blue}{2}}^{\textcolor{green}{1}}\\
    0 & N_{\textcolor{red}{3},\textcolor{blue}{2}}^{\textcolor{green}{2}}
\end{bmatrix}
\begin{bmatrix}
    0 & 0 & h_{\textcolor{red}{2}}^{\textcolor{green}{0}} & d_{\textcolor{red}{3}}^{\textcolor{green}{0}}\\
    0 & 0 & h_{\textcolor{red}{2}}^{\textcolor{green}{1}} & d_{\textcolor{red}{3}}^{\textcolor{green}{1}}
\end{bmatrix}\nonumber\\
&+
\underbrace{
\begin{bmatrix}
    N_{\textcolor{red}{4},\textcolor{blue}{2}}^{\textcolor{green}{0}} & 0\\
    N_{\textcolor{red}{4},\textcolor{blue}{2}}^{\textcolor{green}{1}} & N_{\textcolor{red}{4},\textcolor{blue}{2}}^{\textcolor{green}{0}}\\
    N_{\textcolor{red}{4},\textcolor{blue}{2}}^{\textcolor{green}{2}} & N_{\textcolor{red}{4},\textcolor{blue}{2}}^{\textcolor{green}{1}}\\
    0 & N_{\textcolor{red}{4},\textcolor{blue}{2}}^{\textcolor{green}{2}}
\end{bmatrix}}_{=\mathbf{0}}
\begin{bmatrix}
    0 & 0 & 0 & h_{\textcolor{red}{3}}^{\textcolor{green}{0}}\\
    0 & 0 & 0 & h_{\textcolor{red}{3}}^{\textcolor{green}{1}}
\end{bmatrix}.
\end{align}
The first and last terms in Eq.~\ref{eq:basis matrix in Toeplitz representation} equal to $\mathbf{0}$, because the corresponding basis functions (i.e., $B_{0,2}$ and $B_{4,2}$) are not defined in $[\tau_3, \tau_4]$ (see the triangular computation scheme in \cite{mtuBspline}).

% eq
\begin{align}
\label{eq:recursive basis matrix}
    Eq.~\ref{eq:basis matrix in Toeplitz representation} 
&= 
\begin{bmatrix}
    N_{\textcolor{red}{1},\textcolor{blue}{2}}^{\textcolor{green}{0}}\\
    N_{\textcolor{red}{1},\textcolor{blue}{2}}^{\textcolor{green}{1}}\\
    N_{\textcolor{red}{1},\textcolor{blue}{2}}^{\textcolor{green}{2}}\\
    0
\end{bmatrix}
\begin{bmatrix}
    h_{\textcolor{red}{0}}^{\textcolor{green}{0}} & d_{\textcolor{red}{1}}^{\textcolor{green}{0}} & 0 & 0
\end{bmatrix}
+
\begin{bmatrix}
    0\\
    N_{\textcolor{red}{1},\textcolor{blue}{2}}^{\textcolor{green}{0}}\\
    N_{\textcolor{red}{1},\textcolor{blue}{2}}^{\textcolor{green}{1}}\\
    N_{\textcolor{red}{1},\textcolor{blue}{2}}^{\textcolor{green}{2}}
\end{bmatrix}
\begin{bmatrix}
    h_{\textcolor{red}{0}}^{\textcolor{green}{1}} & d_{\textcolor{red}{1}}^{\textcolor{green}{1}} & 0 & 0
\end{bmatrix}\nonumber\\
&+
\begin{bmatrix}
    N_{\textcolor{red}{2},\textcolor{blue}{2}}^{\textcolor{green}{0}}\\
    N_{\textcolor{red}{2},\textcolor{blue}{2}}^{\textcolor{green}{1}}\\
    N_{\textcolor{red}{2},\textcolor{blue}{2}}^{\textcolor{green}{2}}\\
    0
\end{bmatrix}
\begin{bmatrix}
    0 & h_{\textcolor{red}{1}}^{\textcolor{green}{0}} & d_{\textcolor{red}{2}}^{\textcolor{green}{0}} & 0
\end{bmatrix}
+
\begin{bmatrix}
    0\\
    N_{\textcolor{red}{2},\textcolor{blue}{2}}^{\textcolor{green}{0}}\\
    N_{\textcolor{red}{2},\textcolor{blue}{2}}^{\textcolor{green}{1}}\\
    N_{\textcolor{red}{2},\textcolor{blue}{2}}^{\textcolor{green}{2}}
\end{bmatrix}
\begin{bmatrix}
    0 & h_{\textcolor{red}{1}}^{\textcolor{green}{1}} & d_{\textcolor{red}{2}}^{\textcolor{green}{1}} & 0
\end{bmatrix}\nonumber\\
&+
\begin{bmatrix}
    N_{\textcolor{red}{3},\textcolor{blue}{2}}^{\textcolor{green}{0}}\\
    N_{\textcolor{red}{3},\textcolor{blue}{2}}^{\textcolor{green}{1}}\\
    N_{\textcolor{red}{3},\textcolor{blue}{2}}^{\textcolor{green}{2}}\\
    0
\end{bmatrix}
\begin{bmatrix}
    0 & 0 & h_{\textcolor{red}{2}}^{\textcolor{green}{0}} & d_{\textcolor{red}{3}}^{\textcolor{green}{0}}
\end{bmatrix}
+
\begin{bmatrix}
    0\\
    N_{\textcolor{red}{3},\textcolor{blue}{2}}^{\textcolor{green}{0}}\\
    N_{\textcolor{red}{3},\textcolor{blue}{2}}^{\textcolor{green}{1}}\\
    N_{\textcolor{red}{3},\textcolor{blue}{2}}^{\textcolor{green}{2}}
\end{bmatrix}
\begin{bmatrix}
    0 & 0 & h_{\textcolor{red}{2}}^{\textcolor{green}{1}} & d_{\textcolor{red}{3}}^{\textcolor{green}{1}}
\end{bmatrix}\nonumber\\
&=
%line 2
\begin{bmatrix}
N_{\textcolor{red}{1},\textcolor{blue}{2}}^{\textcolor{green}{0}} & N_{\textcolor{red}{2},\textcolor{blue}{2}}^{\textcolor{green}{0}} & N_{\textcolor{red}{3},\textcolor{blue}{2}}^{\textcolor{green}{0}}\\
N_{\textcolor{red}{1},\textcolor{blue}{2}}^{\textcolor{green}{1}} & N_{\textcolor{red}{2},\textcolor{blue}{2}}^{\textcolor{green}{1}} &     N_{\textcolor{red}{3},\textcolor{blue}{2}}^{\textcolor{green}{1}}\\
N_{\textcolor{red}{1},\textcolor{blue}{2}}^{\textcolor{green}{2}} & N_{\textcolor{red}{2},\textcolor{blue}{2}}^{\textcolor{green}{2}} &     N_{\textcolor{red}{3},\textcolor{blue}{2}}^{\textcolor{green}{2}}\\
0 & 0 & 0
\end{bmatrix}
\begin{bmatrix}
h_{\textcolor{red}{0}}^{\textcolor{green}{0}} & d_{\textcolor{red}{1}}^{\textcolor{green}{0}} & 0 & 0\\
0 & h_{\textcolor{red}{1}}^{\textcolor{green}{0}} & d_{\textcolor{red}{2}}^{\textcolor{green}{0}} & 0\\
0 & 0 & h_{\textcolor{red}{2}}^{\textcolor{green}{0}} & d_{\textcolor{red}{3}}^{\textcolor{green}{0}}
\end{bmatrix}\nonumber\\
&+
\begin{bmatrix}
0 & 0 & 0\\
N_{\textcolor{red}{1},\textcolor{blue}{2}}^{\textcolor{green}{0}} & N_{\textcolor{red}{2},\textcolor{blue}{2}}^{\textcolor{green}{0}} & N_{\textcolor{red}{3},\textcolor{blue}{2}}^{\textcolor{green}{0}}\\
N_{\textcolor{red}{1},\textcolor{blue}{2}}^{\textcolor{green}{1}} & N_{\textcolor{red}{2},\textcolor{blue}{2}}^{\textcolor{green}{1}} &     N_{\textcolor{red}{3},\textcolor{blue}{2}}^{\textcolor{green}{1}}\\
N_{\textcolor{red}{1},\textcolor{blue}{2}}^{\textcolor{green}{2}} & N_{\textcolor{red}{2},\textcolor{blue}{2}}^{\textcolor{green}{2}} &     N_{\textcolor{red}{3},\textcolor{blue}{2}}^{\textcolor{green}{2}}
\end{bmatrix}
\begin{bmatrix}
h_{\textcolor{red}{0}}^{\textcolor{green}{1}} & d_{\textcolor{red}{1}}^{\textcolor{green}{1}} & 0 & 0\\
0 & h_{\textcolor{red}{1}}^{\textcolor{green}{1}} & d_{\textcolor{red}{2}}^{\textcolor{green}{1}} & 0\\
0 & 0 & h_{\textcolor{red}{2}}^{\textcolor{green}{1}} & d_{\textcolor{red}{3}}^{\textcolor{green}{1}}
\end{bmatrix}\nonumber\\
&\stackrel{(Eq.~\ref{eq:B-spline in basis translated representation})}{=}
% line 3
\begin{bmatrix}
\bM^{\textcolor{blue}{2}}(3)\\
\mathbf{0}^{T}
\end{bmatrix}
\begin{bmatrix}
h_{\textcolor{red}{0}}^{\textcolor{green}{0}} & d_{\textcolor{red}{1}}^{\textcolor{green}{0}} & 0 & 0\\
0 & h_{\textcolor{red}{1}}^{\textcolor{green}{0}} & d_{\textcolor{red}{2}}^{\textcolor{green}{0}} & 0\\
0 & 0 & h_{\textcolor{red}{2}}^{\textcolor{green}{0}} & d_{\textcolor{red}{3}}^{\textcolor{green}{0}}
\end{bmatrix}
+
\begin{bmatrix}
\mathbf{0}^{T}\\
\bM^{\textcolor{blue}{2}}(3)
\end{bmatrix}
\begin{bmatrix}
h_{\textcolor{red}{0}}^{\textcolor{green}{1}} & d_{\textcolor{red}{1}}^{\textcolor{green}{1}} & 0 & 0\\
0 & h_{\textcolor{red}{1}}^{\textcolor{green}{1}} & d_{\textcolor{red}{2}}^{\textcolor{green}{1}} & 0\\
0 & 0 & h_{\textcolor{red}{2}}^{\textcolor{green}{1}} & d_{\textcolor{red}{3}}^{\textcolor{green}{1}}
\end{bmatrix}\nonumber\\
&=
% line 4
\begin{bmatrix}
\bM^{\textcolor{blue}{2}}(3)\\
\mathbf{0}^{T}
\end{bmatrix}
\begin{bmatrix}
1-d_{\textcolor{red}{1}}^{\textcolor{green}{0}}& d_{\textcolor{red}{1}}^{\textcolor{green}{0}} & 0 & 0\\
0 & 1-d_{\textcolor{red}{2}}^{\textcolor{green}{0}} & d_{\textcolor{red}{2}}^{\textcolor{green}{0}} & 0\\
0 & 0 & 1-d_{\textcolor{red}{3}}^{\textcolor{green}{0}} & d_{\textcolor{red}{3}}^{\textcolor{green}{0}}
\end{bmatrix}\nonumber\\
&+
\begin{bmatrix}
\mathbf{0}^{T}\\
\bM^{\textcolor{blue}{2}}(3)
\end{bmatrix}
\begin{bmatrix}
-d_{\textcolor{red}{1}}^{\textcolor{green}{1}} & d_{\textcolor{red}{1}}^{\textcolor{green}{1}} & 0 & 0\\
0 & -d_{\textcolor{red}{2}}^{\textcolor{green}{1}} & d_{\textcolor{red}{2}}^{\textcolor{green}{1}} & 0\\
0 & 0 & -d_{\textcolor{red}{3}}^{\textcolor{green}{1}} & d_{\textcolor{red}{3}}^{\textcolor{green}{1}}
\end{bmatrix},
\end{align}
and $\bM^0(3) = B_{3,0}(u) = 1$, where $u = \frac{\tau-\tau_3}{\tau_4-\tau_3}\in[0,1]$.
To understand the second last equation, please recall Eq.~\ref{eq:B-spline in basis translated representation} that the basis matrix $\bM$ is made up by coefficients of the polynomials (basis functions).
To further help memorizing the elements of the basis matrix, please refer to Fig.~\ref{fig:B-spline Triangle Computation Scheme}.
To construct basis matrix $\bM^k(j)$, just look up the column with the corresponding degree (the second subscript) in the blue triangle, and then apply Eq.~\ref{eq:B-spline in basis translated representation}.
% float

Eq.~\ref{eq:recursive basis matrix} can be regarded as the recursive definition of basis matrix.
It can be used in the symbolic computation of NURBS.

%++++++++++++++++++++++++++++++++++++++++++++++++++
\subsection{Basis Matrix $\bM^{k}(j)$ of Uniform B-Spline}
In this section, we provide the general term formula of $\bM^{k}(j)$ for uniform B-spline.
% eq
\begin{align}
\bM^{\textcolor{blue}{k}}(j) &= \frac{1}{k}\Bigg\{ 
\begin{bmatrix}
\bM^{\textcolor{blue}{k-1}}(j)\\
\mathbf{0}^{T}
\end{bmatrix}
\begin{bmatrix}
k+1-j& j-1 & {~} & {~} & 0\\
0 & k+2-j & j-2 & {~} & {~}\\
{~} & {~} & \ddots & \ddots & {~}\\ 
0 & {~} & {~} & k+3-j & 0
\end{bmatrix}\nonumber\\
&+
\begin{bmatrix}
\mathbf{0}^{T}\\
\bM^{\textcolor{blue}{k-1}}(j)
\end{bmatrix}
\begin{bmatrix}
-1 & 1 & {~} & {~} & 0\\
{~} & -1 & 1 & {~} & {~}\\
{~} & {~} & \ddots & \ddots & {~}\\
0 & {~} & {~} & -1 & 1
\end{bmatrix}\Bigg\}\nonumber\\
&\stackrel{(\star\star)}{=}
%line 2
\frac{1}{k}\Bigg\{ 
\begin{bmatrix}
\bM^{\textcolor{blue}{k-1}}(j)\\
\mathbf{0}^{T}
\end{bmatrix}
\begin{bmatrix}
1 & k-1 & {~} & {~} & 0\\
0 & 2 & k-2 & {~} & {~}\\
{~} & {~} & \ddots & \ddots & {~}\\ 
0 & {~} & {~} & 3 & 0
\end{bmatrix}\nonumber\\
&+
\begin{bmatrix}
\mathbf{0}^{T}\\
\bM^{\textcolor{blue}{k-1}}(j)
\end{bmatrix}
\begin{bmatrix}
-1 & 1 & {~} & {~} & 0\\
{~} & -1 & 1 & {~} & {~}\\
{~} & {~} & \ddots & \ddots & {~}\\
0 & {~} & {~} & -1 & 1
\end{bmatrix}\Bigg\},
\end{align}
and $\bM^{0}(j) = 1$.

Note that $(\star\star)$ holds based on the fact that $ j=\frac{\underbrace{k+(k+1)+1}_{\text{\# knots}}}{2}-1 = k$.
Unlike the basis matrices of NURBSs, the basis matrices of uniform B-splines of degree $k$ are independent of $t_j$.
The basis matrices 
% for the 
% case $k=3$.
% Several examples of
% basis matrices 
for uniform B-splines are given as follows:
%eq
\begin{align}
\label{eq:constant basis matrix}
\bM^{0}(j) &= 1\nonumber,\\
\bM^{1}(j) &=
\begin{bmatrix}
1 & 0\\
-1 & 1
\end{bmatrix}\nonumber,\\
\bM^{2}(j) &= \frac{1}{2!}
\begin{bmatrix}
1 & 1 & 0\\
-2 & 2 & 0\\
1 & -2 & 1
\end{bmatrix}\nonumber,\\
\bM^{3}(j) &= \frac{1}{3!}
\begin{bmatrix}
1 & 4 & 1 & 0\\
-3 & 0 & 3 & 0\\
3 & -6 & 3 & 0\\
-1 & 3 & -3 & 1
\end{bmatrix}\\
& \vdots\nonumber
\end{align}
There is no need to memorize Eq.(8) in \cite{qin2000general}.
%++++++++++++++++++++++++++++++++++++++++++++
\subsection{Basis Matrices in the Cumulative Formula}
For the cumulative formula (Eq.~\ref{eq:cumulative formula}), we can obtain a similar representation.
Here we still use the specific case ($\textcolor{blue}{k=3}$) as an example.
% eq
\begin{align}
\label{eq:differential cumulative formula}
\bP(u) &= \tilde{B}_{0,\textcolor{blue}{3}}(u)\bP_0 + \sum_{i=1}^{3}\tilde{B}_{i,\textcolor{blue}{3}}(u)\underbrace{(\bP_i - \bP_{i-1})}_{\bd_i},
\end{align}
where $\tilde{B}_{i,k} = \sum_{s=i}^{\textcolor{blue}{3}} B_{s,\textcolor{blue}{3}}(u)$.
Specifically,
% eq
\begin{align}
    \tilde{B}_{0,3} &= B_{0,3} + B_{1,3} + B_{2,3} + B_{3,3}\nonumber\\
    \tilde{B}_{1,3} &= B_{1,3} + B_{2,3} + B_{3,3}\nonumber\\
    \tilde{B}_{2,3} &= B_{2,3} + B_{3,3}\nonumber\\
    \tilde{B}_{3,3} &= B_{3,3}.
\end{align}

Following the format in Eq.~\ref{eq:B-spline in basis translated representation}, the differential cumulative formula (Eq.~\ref{eq:differential cumulative formula}) can be represented, by dropping off $u$ for simplicity, as
% eq
\begin{align}
\bP(u)^T &= [\tilde{B}_{\textcolor{red}{0},3}~\tilde{B}_{\textcolor{red}{1},3}~\tilde{B}_{\textcolor{red}{2},3}~\tilde{B}_{\textcolor{red}{3},3}]
\begin{bmatrix}
\bP_{\textcolor{red}{0}}^T\\
\bd_{\textcolor{red}{1}}^T\\
\bd_{\textcolor{red}{2}}^T\\
\bd_{\textcolor{red}{3}}^T
\end{bmatrix}\nonumber\\
&=
% line 2
\Bigg\{[B_{0,3}~B_{1,3}~B_{2,3}~B_{3,3}] + [B_{1,3}~B_{2,3}~B_{3,3}~0] + \cdots\nonumber\\
&+ [B_{2,3}~B_{3,3}~0~0] + [B_{3,3}~0~0~0]
\Bigg\}
\begin{bmatrix}
\bP_{\textcolor{red}{0}}^T\\
\bd_{\textcolor{red}{1}}^T\\
\bd_{\textcolor{red}{2}}^T\\
\bd_{\textcolor{red}{3}}^T
\end{bmatrix}\nonumber\\
&=
% line 3
[1~u~u^2~u^3]
\Bigg\{
[\bm_0~\bm_1~\bm_2~\bm_3] + [\bm_1~\bm_2~\bm_3~0]\nonumber\\
&+
[\bm_2~\bm_3~0~0] + [\bm_3~0~0~0] 
\Bigg\}
\begin{bmatrix}
\bP_{\textcolor{red}{0}}^T\\
\bd_{\textcolor{red}{1}}^T\\
\bd_{\textcolor{red}{2}}^T\\
\bd_{\textcolor{red}{3}}^T
\end{bmatrix}\nonumber\\
&=
% line 4
[1~u~u^2~u^3]\cdot
\begin{bmatrix}
\sum_{s=0}^{3}\bm_s & \vert & \sum_{s=1}^{3}\bm_s & \vert & \sum_{s=2}^{3}\bm_s & \vert & \bm_3
\end{bmatrix}
\begin{bmatrix}
\bP_{\textcolor{red}{0}}^T\\
\bd_{\textcolor{red}{1}}^T\\
\bd_{\textcolor{red}{2}}^T\\
\bd_{\textcolor{red}{3}}^T
\end{bmatrix}\nonumber\\
% line 5
&\stackrel{(\text{substitute Eq.~\ref{eq:constant basis matrix}})}{=}
\frac{1}{3!}[1~u~u^2~u^3]\cdot
\begin{bmatrix}
6 & 5 & 1 & 0\\
0 & 3 & 3 & 0\\
0 & -3 & 3 & 0\\
0 & 1 & -2 & 1
\end{bmatrix}
\begin{bmatrix}
\bP_{\textcolor{red}{0}}^T\\
\bd_{\textcolor{red}{1}}^T\\
\bd_{\textcolor{red}{2}}^T\\
\bd_{\textcolor{red}{3}}^T
\end{bmatrix}\nonumber\\
% line 6
&=
\underbrace{\begin{bmatrix}
1 & \frac{5+3u-3u^2+u^3}{6} & \frac{1+3u+3u^2-2u^3}{6} & \frac{u^3}{6}
\end{bmatrix}}_{\boldsymbol{\lambda}}
\begin{bmatrix}
\bP_{\textcolor{red}{0}}^T\\
\bd_{\textcolor{red}{1}}^T\\
\bd_{\textcolor{red}{2}}^T\\
\bd_{\textcolor{red}{3}}^T
\end{bmatrix}\nonumber\\
% line 7
&=
\bP_{\textcolor{red}{0}}^T + \sum_{i=1}^{3}\boldsymbol{\lambda}_{\textcolor{red}{i}}(u)\bd_{\textcolor{red}{i}}^T,
\end{align}
where $\blambda_0 = 1$, $u = \frac{\tau - \tau_3}{\tau_4-\tau_3}\in[0,1]$.
\section{FAQs}

\begin{itemize}
    \item Q1: \textbf{Knots} vs \textbf{Control Points}\\
    A: Knots are a list of positions in the parametric domain (i.e., $\tau_i \in [0,1]$.)
    For uniform B-splines, knots are evenly distributed in the parametric domain.
    The number of knots is determined if the degree of B-spline is known (see Sec.\ref{sec:cox-de boor formula}).
    Control points are design parameters from human's input.
    Once the degree $k$ is set and the control points are provided, one can evaluate the value at any given position $\tau$ in the non-zero domain, which is spanned by the two knots in the middle (e.g., $[\tau_3,\tau_4]$ for a B-spline of degree $3$).
    
    Some papers, such as \cite{sommer2020efficient}, somehow treat knots and control points identically.
    This is not consistent with the majority of the literature.
    Thus, we regard that in \cite{sommer2020efficient} as an improper (wrong) definition; better not use it.
    
    \item Q2: How to understand \textbf{basis translation}?\\
    This is actually a trivial operation (see Eq.~\ref{eq: basis translation}).
    However, I felt confused when I read the descriptions in academic papers (e.g., the $2_{\text{nd}}$ paragraph in Section 4.2 of \cite{sommer2020efficient}, and in Sec. IV of \cite{ding2019efficient}.)
    The confusion is caused mainly by the inconsistent symbolic definition and descriptions.
    In general, the \textbf{basis translation} just simply translates and re-scales the non-zero domain to $[0,1]$ such that the numerical stability is preserved. 
\end{itemize}

\bibliographystyle{plain}
\bibliography{references}
\end{document}